# SSR: A Stall Scheme Reducing Bubbles in Load-Use Hazard of RISC-V Pipeline

Dongchu Su[*], Yong Li[*], Bo Yuan[*]
[*]College of Computer Science, National University of Defense Technology, 410073, Changsha, P.R.China
Email: [*]sudongchu@nudt.edu.cn, [*]yongli@nudt.edu.cn, [*]yuanbo18@nudt.edu.cn

**Abstract.** Modern processors usually adopt pipeline structure and often load data from memory. At that point, the load-use hazard will inevitably occur, which usually stall the pipeline and reduce performance. This paper introduces and compares two schemes to solve load-use hazard. One is the traditional scheme that detect hazard between ID stage and EXE stage, which stalls the pipeline and insert bubbles between the two instructions. In the scheme we proposed, we add a simple bypass unit between EXE and MEM stage that disables the stall of load-use hazard caused by the traditional scheme, which can reduce the bubble between the two instructions. It's quite a considerable benefit in eliminating bubbles especially in the long pipeline or programs of plenty load instructions. The scheme was implemented in the open source RISC-V SoC generator Rocket-chip and synthesized in SMIC 130-nm technology. The results show that the performance of the latter scheme is increased by 6.9% in the Dhrystone benchmark with the reasonable cost of area and power.

## 1 Introduction

There always exiting load-use hazard in processors since the pipeline was proposed, which generally refers to the current operand of the instruction is not the latest. The reason causing load-use hazard is the data dependency which means that the later instructions needs the latest result of the former load instruction while it can't be known in a clock cycle without some special bypass units.

The load-use hazard can affect the correctness of program results and even leading wrong place where program may jump. In the most case, Load-use hazard can cause the program to crash. Therefore, resolving data hazard is a necessary condition to ensure the normal behavior of the program.

The load-use hazard is encountered in every program. The traditional scheme is a stall pipeline for one cycle [2], but it will bring an additional bubble. Paper [1] provide software method to alleviate this problem, which reorders the instructions, but just avoid this situation and cannot completely solve it. Once it occurs, stall and bubbles will inevitably occur. To avoid this stall and bubble, we propose a new hardware micro-architecture called SSR to solve this problem, which was inspired in designing a 5-stage pipeline RISC-V core NF5 [6]. Readers will understand the data-related theory and the detail of high-performance core implementation after reading this paper.

The organization of this paper is as follows: Section 2 is a background introduction, which describes a specific example of RAW hazard and bypass unit in the pipeline, Section 3 describes the nature of the load-use hazard problem and its difference with other RAW with analysis of the shortcomings of traditional stall scheme. Section 4 proposes a new micro-architecture scheme to addresses the load-use hazard and its pseudocode. Section 5 describes the implementation of this micro-architecture on the open source RISC-V SoC generator Rocket-chip [7], and evaluates the performance of the implemented micro-architecture

## 2  Background

**RAW.** There are 4 types of data hazard in processor pipeline, called Read After Write (RAW), Write After Read (WAR), Read After Read (RAR), and Write After Write (WAW). RAW refers to that one instruction read a register after the other has modified it. WAR, RAR and WAW will not appear in the sequential pipeline. Therefore, we mainly discuss the RAW of which load-use hazard is special one.

Figure 1 shows a simple RISC-V 5-stage pipeline which was divided into 5 stages: Instruction Fetch (IF), Instruction Decode (ID), Execute (EXE), Mem Access (MEM), Write Back (WB). The pipeline also shows a RAW situation. The two add instructions update the a1 and a4 registers respectively, but updating the a4 register requires reading the value of the a1 register while the previous add instruction has not written the newest a1 back to the register file. If there is no auxiliary measure, then the a1 read by the latter instruction in ID would be wrong.

**Hazard Detector.** The common method for solving RAW is bypass, which uses hazard detector to solve data hazard. It is a method commonly used in pipelines. Once data hazard occurs, just wait for the latest data to appear and transfer the data to where it needed. As shown in Figure 2, hazard detector can bypass the latest a1 as soon as EXE finishes executing.

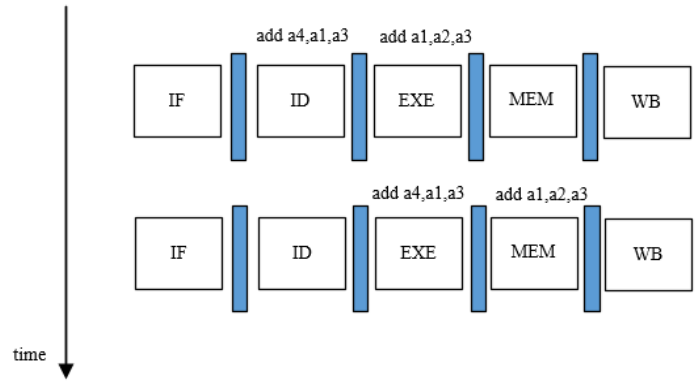

**Fig. 1.** Two RISC-V assembly instruction with RAW in pipeline.

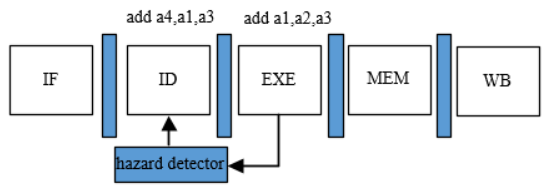

**Fig. 2.** Hazard detector for solving RAW.

## 3    Comprehensive Analysis on Load-use Hazard

The load-use hazard is a special case of RAW, its difference with the RAW hazard generated by the add instruction is that the load-use hazard generated by the load instruction can only bypass data after getting loaded data in the MEM station as shown in Figure 3. The lw instruction must reach the MEM station first, and when the lw reaches the MEM station, the beqz instruction has arrived at the EXE station. If lw instruction has not completed the data loading, beqz instruction cannot judge the jump, otherwise may cause unexpected jumps.

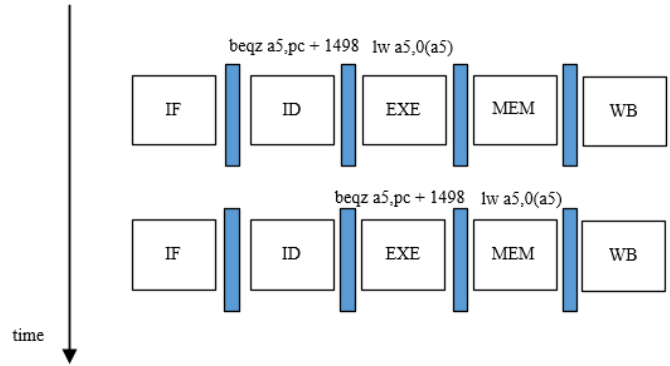

**Fig. 3.** Hazard detector for solving RAW.

As shown in Figure 4, the traditional stall scheme in [1], [2], [3] is to stall beqz instruction and insert a bubble (nop instruction) between the two hazard instructions, so that the hazard detector located at the ID station can detect the data hazard between ID and MEM. It can be concluded that compared to the data hazard caused by the add instruction, the load-use hazard requires the stall mechanism, and will bring extra bubbles with the traditional stall scheme.

Software compilers can use the insertion of irrelevant instructions and reordering instructions to alleviate the load-use problem [1], but just avoid this situation and cannot completely solve it. Once it occurs, stall and bubbles will inevitably occur, because the source of the problem is The hazard detector located at the ID station cannot detect the hazard between EXE and MEM.

However, if the hazard detector is completely placed in the EXE, another problem occurs, that is, it cannot cope with the case where the WB instruction writes a register when the ID instruction needs to read the it, that is, the detection unit of the EXE cannot detect the data hazard between the ID and the WB. Unless you add another read port to the register file and bypass to EXE in proper occasion, this is obviously not cost-effective. Therefore, we proposed a new micro-architecture of this scheme for the load-use hazard in the next section.

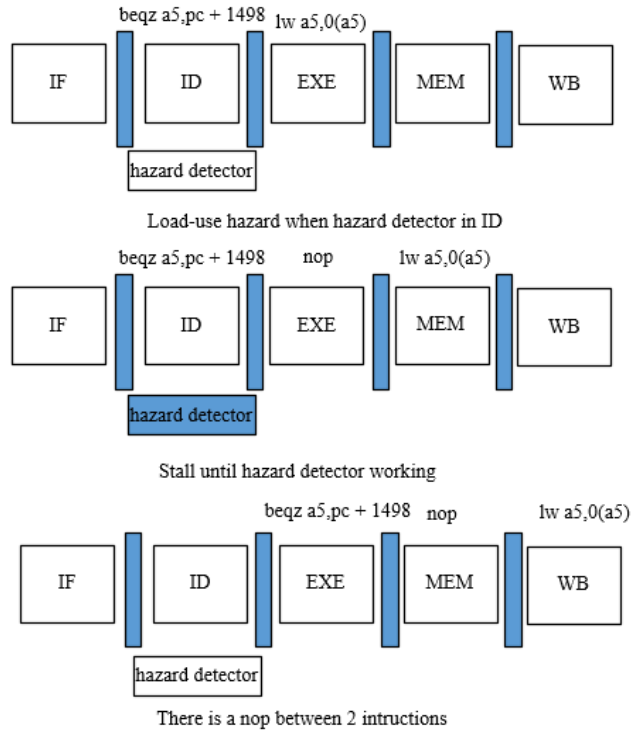

**Fig. 4.** Traditional stall scheme for load-use hazard.

## 4  SSR Micro Architecture

Figure 5 shows the SSR micro-architecture we proposed to solve the load-use hazard. Based on the original scheme that all hazard detector in ID station, a small ld hazard detector unit is added to the EXE station to detect the data hazard between EXE and MEM. The correct data can be transferred to EXE through a combinational logic block. Even if a cache miss stall occurs in MEM station, the ld hazard detector unit can work normally after the pipeline is restored. It can be proved that this architecture is applicable to pipelines of any number of stages. Algorithm 1 is pseudo code that implements the ld hazard detector unit.

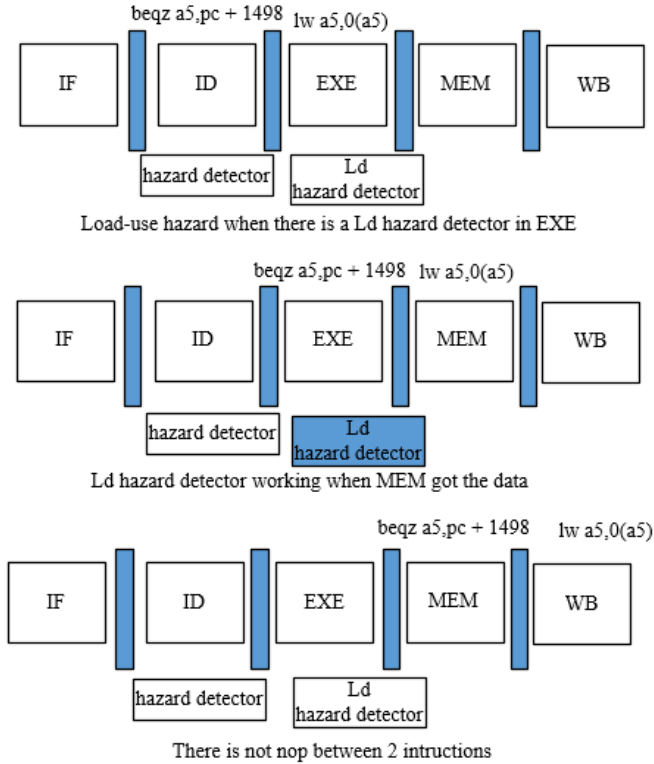

**Fig. 5.** A new micro-architecture to solve load-use hazard.

| Algorithm 1 | Pseudocode for ld hazard detector unit |
|---|---|

if(IDEX.Rs != 0 && EXMEM.RdWrtEn && IDEX.Rs == EXMEM.Rd && IsLoadInstruction)
  Then bypass from EXMEM
  else if(IDEX.Rs != 0 && EXMEM.RdWrtEn && IDEX.Rs == EXMEM.Rd && IsLoadInstruction==0)
  Then bypass from MEM
  else
  do nothing

## 5  Evaluation of SSR

We implemented this micro-architecture on the open source RISC-V SoC generator Rocket-chip [4], [7]. Rocket-chip is a highly parameterized SoC generator implemented by Chisel [6] released by UCB. Rocket-chip has been tape-out for many times and is capable of running the Linux operating system. The Rocket-chip SoC generator can generate C++ simulators and Verilog implementations equivalent to SoC designs.

We configure the Rocket-chip generator as the DefaultConfig and use Dhrystone to evaluate the proposed micro-architecture. The Rocket core's bypass scheme is also the traditional stall scheme. We made the following changes to Rocket core : id_load_use signal was deleted firstly by setting bool false of its value, and the second was to add a detector to detect the load-use hazard between EXE and MEM, and add it to the source of bypass for selection at the ALU input.

To verify the performance of the SSR, we use a cycle-accurate C++ emulator generated by Rocket-chip generator to perform evaluation and compare it with the traditional stall scheme. We then used the Synopsys Design Compiler tools to synthesize the Verilog file which also generated by Rocket-chip generator for evaluating the of area and power consumption. We summarized the performance speed up and the overhead comparison messages in Table 1.

The main reason for the increase in area and power consumption is that the implementation of SSR uses 2 chisel Mux units, which may be too redundant and can be optimized. And according to statistics, about 10% of Dhrystone instructions will encounter a load-use hazard, but the optimization Up to 6.9%, this may be due to the cache miss related to Rocket, resulting in multiple stalls. In this case, the SSR's optimization capability is limited.

**Table 1.** Performance and overhead comparison summary.

| Technology | SMIC 130nm | | |
|---|---|---|---|
| Compared schemes | Stall Scheme | SSR | Optimization rate |
| Supply voltage | 1.2V | 1.2V | - |
| Clock rate | 500Mhz | 500Mhz | - |
| Dynamic power consumption | 72.6431 mW | 78.7678 mW | -8.4% |
| Core area | 0.344171 mm$^2$ | 0.366288 mm$^2$ | -6.4% |
| Clocks spent by benchmark | 247888 | 231862 | +6.9% |
| Microseconds for one run through Dhrystone | 495us | 463us | +6.9% |
| Dhrystones per second | 2017 | 2153 | +6.7% |

## 6 Conclusion

In this paper, we proposed a new micro-architecture called SSR for solving load-use hazard, as well as getting rid of the unnecessary bubbles and stall of the traditional stall scheme. We implemented this micro-architecture in the open source RISC-V SoC generator Rocket-chip and indeed improved the performance at an acceptable level of power consumption and area consumption, making contribution to the open source community and the research of high performance computing.

In the Future, our work will focus on the exploration of lower power consumption and area design while keeping the performance basically unchanged. At present implementation, there are two added Chisel Mux exiting before the ex_rs signal, which results in a relatively large area and power consumption overhead, among which there is a lot of room for optimization.

**Acknowledgments.** This research work is fully supported by NUDT-NF5 RISC-V development team.